\documentclass{article}
\usepackage{amsmath}
\usepackage{graphicx}

\newtheorem{theorem}{Theorem}
\newtheorem{acknowledgement}[theorem]{Acknowledgement}

\begin{document}

\author{Fabrizio Canfora and Gaetano Vilasi \thanks{%
Corresponding author: Gaetano Vilasi, Universit\`{a} di Salerno,
Dipartimento di Fisica ''E.R.Caianiello'', Via S.Allende, 84081 Baronissi
(Salerno) Italy; e-mail: vilasi@sa.infn.it, Phone: +39-089-965317, Fax:
+39-089-965275.} \\
\textit{Istituto Nazionale di Fisica Nucleare, GC di Salerno, Italy. }\\
\textit{Dipartimento di Fisica ''E.R.Caianiello'', Universit\`{a} di Salerno.%
}}
\title{PP-waves, Israel's Matching conditions, Brane-world scenarios and BPS states
in gravity}
\date{}
\maketitle

\begin{abstract}
The matching between two 4-dimensional PP-waves is discussed by using
Israel's matching conditions. Physical consequences on the dynamics of
(cosmic) strings are analyzed. The extension to space-time of arbitrary
dimension is discussed and some interesting features related to the brane
world scenario, BPS states in gravity and Dirac-like quantization conditions
are briefly described.

\end{abstract}

\section{Introduction}

\noindent Because of huge observational efforts devoted to the detection of
gravitational waves, there is a growing interest in the theoretical analysis
of exact wave-like solutions of Einstein equations. Indeed, although the
linearized theory has played a prominent role in clarifying some general
features related to the propagation and to the detection of gravitational
waves, the linearization process could hide interesting phenomena that only
the exact theory may disclose \cite{Ch91}. Moreover, in order to have a good
calibration of detectors, it is important to know in advance, with high
precision, the range of frequencies.

Among all known exact gravitational waves, plane fronted gravitational waves
with parallel rays (PP-waves), first introduced by Peres \cite{Pe59, Pe60},
are particularly interesting both from the theoretical and the experimental
point of view. They are space-times $g$\ admitting a covariantly constant
isotropic Killing vector field and a special class is completely
characterized \cite{SVV00} by the invariance for the \textit{non Abelian
two-dimensional Lie algebra}\footnote{%
Recall that, up to isomorphisms, there are only two $2-$dimensional Lie
algebras: Abelian and non-Abelian, the corresponding generators having the
canonical commutation relation $\left[ X,\,\,\,Y\right] =sY,$ with $s$ =0,1
respectively.} generated by the Killing vector fields $X$ and $Y$ such that:

\begin{itemize}
\item[\textit{i)}]  their commutator is isotropic: $g\left( Y,Y\right) =0;$

\item[\textit{ii)}]  the distribution $\mathcal{D}^{\bot }$, orthogonal to
the distribution$\mathcal{\ D}$ generated by $X$ and $Y$, $\ $is integrable
and transversal to $\mathcal{D}$.
\end{itemize}

From theoretical side, asymptotically flat PP-waves have a longitudinal
polarization \cite{Pe59, Pe60, CVV02a, CVV02b}, and are very naturally
coupled to cosmic strings. Moreover, supergravity and superstring theory on
a PP-waves background have very interesting properties; in particular, some
superstring theories on non singular PP-waves background are not so far from
being exactly quantizable: to be more precise, the spectrum of closed string
theory on such background is known for $p^{+}\neq0$ but without the
inclusion of interactions (see, for example, \cite{BMN02, Me02, IKM02, RT02}
and references therein). The physics of \textit{D-branes} on PP-waves
backgrounds also has interesting properties which, in some cases, can be
studied analytically \cite{Ali02, OPS03, PS03}. Furthermore, it is not
unlikely that, once experimental devices will detect gravitational waves, it
will be possible to distinguish the experimental signature of a PP-wave from
the one of other gravitational waves \cite{CV03}.

In this paper we will study the matching between two PP-waves moving in
opposite spatial directions. Because of the linear superposition law of
PP-waves \cite{Za73, Gri91, CVV02a, CVV02b}, two PP-waves travelling in the
same direction do not interact, but simply superpose. Moreover, being
(cosmic) strings and PP-waves strictly related \cite{CV03}, this analysis
turns out to be a natural theoretical arena to understand the highly non
trivial physical processes occurring, for instance, in the early universe in
which cosmic strings and other topological defects, such as domain walls,
naturally occur.

The paper is organized in the following way: in the first section some
properties of PP-waves that will be useful to suggest a reasonable form for
the metric are shortly described. Then, Israel's matching conditions are
introduced. In the third section the matching between two PP-waves
travelling in opposite spatial direction are analyzed and the physical
consequences on the dynamics of cosmic strings and $\Sigma$ are drawn. The
extension to space-times of arbitrary dimensions is discussed and some
interesting features related to the \textit{brane world scenarios} (see, for
example, \cite{HW96a, Ar98a, Ar98b, RS99a, RS99b} and references therein)
are briefly described. Particular emphasis has been given to the
five-dimensional case in connection with the role of BPS states in gravity%
\footnote{%
Such as the Majumdar-Papapetrou solutions which, as it will be explained
below, are naturally related to PP-waves.}, and the relation between the
Dirac quantization condition and the mass quantization of extremal charged
black-holes.

\section{The metric}

PP-waves provide interesting non trivial backgrounds for superstring theory.
In this section we will study the matching between two of them.

There exist Impulsive PP-waves which are usually described either by a
distributional space-time metric or, alternatively, by a continuous one
whose curvature tensor involves distributions. Moreover, for impulsive
PP-waves constructed by gluing together two copies of Minkowski space along
a null hypersurface, according to Penrose's scissors \cite{Pe72}, a C1-atlas
exists \cite{CD87} in which the metric components are continuous. A
discussion on the Penrose junction conditions can be found in \cite{KS99}.

However, the PP-waves that string theorists usually consider (see, for
example, \cite{BMN02} \cite{Ali02} \cite{Me02} \cite{IKM02} \cite{RT02}) are
non singular: this means that the harmonic function which parametrizes such
waves is assumed to be quadratic. Instead, we will consider asymptotically
flat PP-waves, that is, PP-waves whose harmonic function has a finite number
of $\delta-$like singularities. Such PP-waves space-times, as we will now
explain, have a finite number of string-like singularities. These
string-like singularities have very interesting dynamical properties that
can be partly analyzed by studying the matching of singular PP-waves.

Let us start by writing the metric representing a PP-wave propagating in the
positive $z-$direction:
\begin{equation}
g^{+}=2dudv+H^{+}(x,y,u)du^{2}-dx^{2}-dy^{2},  \label{colright}
\end{equation}
where $z=\left( u-v\right) /2$, and the vacuum field equations imply
\begin{equation}
\left( \partial_{x}^{2}+\partial_{y}^{2}\right) H^{+}=0\text{.}
\label{prima}
\end{equation}
Because of Eq. (\ref{prima}), coordinates $(u,v,x,y)$ are harmonic. However,
in order to have spatially asymptotically flat gravitational fields (a
finite number of) $\delta-$like singularities in the $x-y$ plane (that is,
by taking into account the third spatial dimension, string-like
singularities) have to be allowed \cite{CV03}, so that $H^{+}$ fulfils the
following equation:
\begin{equation}
\left( \partial_{x}^{2}+\partial_{y}^{2}\right)
H^{+}=\sum_{i}\mu_{i}^{+}\delta^{(2)}\left( x_{i}(u),y_{i}(u)\right) ,
\label{la1}
\end{equation}
where $\mu_{i}^{+}$\ represents the energy per unit of length of the $i-$th
cosmic strings (see, for example, \cite{VS00} and references therein) and
the positions of the singularities, in principle, could change with $u$. It
is also worth to note here that, besides to be exact solutions of Einstein
equations, PP-waves are solutions of the linearized field equations on flat
background too \cite{CVV02a} \cite{CVV02b}, this allowing the analysis of
the polarization by standard methods, \cite{We72} \cite{Wa84}. Thus, cosmic
strings are naturally coupled to asymptotically flat PP-waves. Of course, an
asymptotically flat PP-wave propagating in the negative $z-$direction can be
written in the following form:
\begin{align}
g^{-} & =2dudv+H^{-}(x,y,v)dv^{2}-dx^{2}-dy^{2}  \label{coleft} \\
\left( \partial_{x}^{2}+\partial_{y}^{2}\right) H^{-} & =\sum_{i}\mu
_{i}^{-}\delta^{(2)}\left( x_{i}(v),y_{i}(v)\right) ,  \label{la2}
\end{align}
and previous comments also apply in this case.

In this paper, we will study a space-time in which there are PP-waves
travelling in opposite spatial direction. Of course, it is very hard to find
the exact analytical form of a space-time of this kind. However, the
Israel's procedure gives interesting analytical conditions constraining the
''coexistence'' of different kinds of PP-waves. Such conditions, as we will
see in the following, have a natural interpretation as topological
conservation laws.

From the ''scattering-theoretic'' point of view, a natural ansatz for a
metric describing the collision of two PP-waves travelling in opposite
spatial direction is the following:
\begin{equation}
g=2dudv+H^{-}(x,y,u,v)dv^{2}+H^{+}(x,y,u,v)du^{2}-dx^{2}-dy^{2},
\label{col1}
\end{equation}
and the natural boundary conditions, ensuring that asymptotically (that is,
when the two waves are very far apart) these two PP-waves do not interact,
are
\begin{equation}
\lim_{u\rightarrow+\infty}\partial_{v}H^{+}=\lim_{v\rightarrow+\infty}%
\partial_{u}H^{-}=\lim_{u\rightarrow+\infty}H^{-}=\lim_{v\rightarrow+\infty
}H^{+}=0.  \label{col2}
\end{equation}

Thus, previous conditions ensure that, metric (\ref{col1}) asymptotically
reduces to the ''superposition'' of gravitational fields (\ref{colright})
and (\ref{coleft}). It can be shown that vacuum Einstein equations admit non
trivial solutions for the metric (\ref{col1}) with the boundary conditions (%
\ref{col2}). Since in the general case there are no Killing fields, the
field equations are extremely complicated and no simple analytic expression
can be found for $H^{+}$\ and $H^{-}$. However, many analytic informations
can be obtained by using the Israel's matching conditions \cite{Is66}.
Roughly speaking, in our case the Israel method gives the conditions to
match a PP-wave propagating in the positive $z-$direction with a PP-wave
propagating in the negative $z-$direction on the 3-dimensional\footnote{%
Of course, in general $N$-dimensional space-time, such a timelike
hypersurface would be ($N-1)$-dimensional.} time-like hypersurface $2z=u-v=0$%
. In other words, in an $N$-dimensional space-time, the scattering region
between two colliding waves can be represented by an ($N-1)$-dimensional
timelike hypersurface with non trivial properties and the Israel matching
conditions can be thought, in such cases, as dynamical constraints on the
scattering process. At a first glance, this seems a very rough approximation
of a collision process. In fact, as it is well known (see, for example, \cite
{Gri91}), the collision takes places at $u=0$ and $v=0$ and affects the
whole causal future of the interaction region itself. By using the Israel
matching conditions, we let ''collapse'' (maybe, it would be better to say
''we project'') the whole causal future of $u=0$ and $v=0$ on the
hypersurface $\Sigma\equiv\left\{ u=v\right\} $. However, as we will show
below, our main results are topological in nature and so they survive to
smooth deformations such as the above ''projection''.

\section{The Israel Matching conditions}

Let us consider a $(N-1)-$dimensional non null hypersurface $\Sigma$
embedded in a $N-$dimensional Lorentzian manifold $M$ fulfilling the
Einstein equations. The action will be the sum of the standard bulk term
(with contributions from the matters field also) and the boundary term:
\begin{align*}
S & =S^{B}+S^{\Sigma} \\
S^{B} & =\frac{1}{2\kappa^{2}}\int_{M}d^{N}x\sqrt{-g}\left(
R+L_{m}^{B}\right) \\
S^{\Sigma} & =-\frac{1}{\kappa^{2}}\int_{\Sigma^{\pm}}d^{N-1}x\sqrt {-h}K-%
\frac{1}{\kappa^{2}}\int_{\Sigma^{\pm}}d^{N-1}x\sqrt{-h}L_{m}^{\Sigma^{\pm}},
\end{align*}
where $h_{ab}$ is the induced metric on $\Sigma$, $K$ is the trace of the
extrinsic curvature $K_{ab}$, $\Sigma^{\pm}$ stands for fields living on the
right or on the left part of $\Sigma$, $L_{m}^{B}$ is the bulk Lagrangian
density for the matter fields and $L_{m}^{\Sigma^{\pm}}$ is the boundary
Lagrangian density for matter fields. Then, starting with a right
propagating $g^{+}$ and a left propagating $g^{-}$ PP-waves solutions of
Einstein equations, namely $g^{+}$ and $g^{-}$ respectively, we can match
them to obtain a global (piecewise $C^{2}$) solution provided the following
conditions are fulfilled \cite{Is66}:
\begin{align}
\lim_{x\rightarrow\Sigma^{+}}g^{+}(x) &
=\lim_{x\rightarrow\Sigma^{-}}g^{-}(x),  \label{col3} \\
2\left\langle K_{ab}-h_{ab}K\right\rangle & =-\kappa^{2}S_{ab},  \label{col4}
\\
S_{ab} & =2\frac{\partial L_{mat}^{\Sigma}}{\partial h^{ab}}%
-h_{ab}L_{mat}^{\Sigma},  \label{col5}
\end{align}
where $\left\langle X\right\rangle =\left[ X(\Sigma^{+})-X(\Sigma ^{-})%
\right] /2$ denotes the jump of $X$ between the two sides ($\Sigma^{\pm }$)
of the hypersurface and $S_{ab}$ can be naturally interpreted as the
energy-momentum tensor of the hypersurface \cite{Is66}.

In our case, it is natural to match gravitational fields (\ref{colright})
and (\ref{coleft}) on the timelike hypersurface $u=v$ that can be thought as
the hypersurface $z=0$, $z$ being the propagation direction of the two waves
(see, for example, \cite{CVV02b}). We will first try to solve the Israel
conditions in the vacuum, that is for $S_{ab}=0$. Let us use the induced
local coordinates on the hypersurface
\begin{equation*}
\left( t,x,y\right) \equiv\left( 1,2,3\right)
\end{equation*}
being $2t=u+v$. An explicit computation of the left hand side of Eq. (\ref
{col4}) gives
\begin{align}
\left\langle K_{tx}-h_{tx}K\right\rangle &
=0\Rightarrow\partial_{x}H^{+}=\partial_{x}H^{-}  \label{col6} \\
\left\langle K_{ty}-h_{ty}K\right\rangle &
=0\Rightarrow\partial_{y}H^{+}=\partial_{y}H^{-}.  \label{col7}
\end{align}
Before analyzing the above conditions\footnote{%
That always can be fulfilled either in vacuum with $S_{ab}=0$ or with $%
S_{ab}\neq0$.}, let us give a closer look to condition (\ref{col3}). It is
easy to see that in the vacuum case, that is with $S_{ab}=0$, condition (\ref
{col3}) admits only the trivial solution. Indeed, condition (\ref{col3})
implies
\begin{equation*}
\lim_{u\rightarrow v^{+}}H^{+}=\lim_{v\rightarrow u^{+}}H^{+}=0,
\end{equation*}
so that $H^{+}$ has to depend also on $v$ while $H^{-}$ has to depend also
on $u$, since
\begin{equation*}
\lim_{u\rightarrow v^{+}}H^{+}=0\Rightarrow\lim_{u\rightarrow v^{+}}\frac{%
H^{+}}{\left( u-v\right) ^{\beta}}=1,\quad\beta>0;
\end{equation*}
a similar argument also holds for $H^{-}$. However, according to our ansatz $%
H^{+}$ does not depend on $v$ and $H^{-}$ does not depend on $u$.

To understand the physical meaning of this, it is useful to slight modify
two metrics on the two sides of $\Sigma$\ by replacing $H^{\pm}$ with $%
f(u-v)H^{\pm}$, where $f(u-v)$\ is an even function such that
\begin{equation}
\lim_{x\rightarrow0}f\left( x\right) =0,\quad\lim_{x\rightarrow+\infty
}f\left( x\right) =1.  \label{brana1}
\end{equation}
For example, one could consider
\begin{equation*}
f(x)=\exp\left[ -\frac{L^{2}}{(u-v)^{2}}\right] .
\end{equation*}
In this way, to within few $L$ lengths in the $z-$ direction away from $%
\Sigma$, one recovers the gravitational field of two non-interacting
PP-waves, while on $\Sigma$ (once conditions (\ref{col6}) and (\ref{col7})
have been fulfilled) condition (\ref{col3}) is satisfied because of Eq.(\ref
{brana1}). Finally, we will let $L$ tend to zero to understand what really
happens to the matching hypersurface $\Sigma$. It is easy to check that
Israel matching conditions cannot be satisfied in vacuum anymore although,
to compensate this ''mismatching'', it is sufficient to consider an
energy-momentum tensor $S_{ab}$ whose only non-vanishing component is,
\begin{equation*}
S_{tt}\sim-\left[ 3\partial_{z}f\left( H^{+}-H^{-}\right) -f\partial
_{z}f\left( \left( H^{+}\right) ^{2}-\left( H^{-}\right) ^{2}\right) \right]
,
\end{equation*}
the one with two index along the time direction. Such an energy-momentum
tensor for $\Sigma$\ is non vanishing only$\ $in a region of size of order $%
2L$ in the $z$-direction away from $\Sigma$, so, in the limit $L\rightarrow
0 $, we obtain a distributional energy-momentum (that is, proportional to $%
\delta(z)$) whose support is on $\Sigma$. Such an energy-momentum tensor
represents the energy-momentum tensor of a tensionless membrane placed at $%
z=0$: with respect to the energy-momentum tensor of a domain wall (see, for
example, \cite{VS00} \cite{Sa02}), such an $S_{ab}$ does not have the
spatial components $S_{ij}$ which are usually related to various types of
instabilities.

Thus, cosmic strings naturally end up on membranes and, at the same time,
membranes allow topological interactions between cosmic strings. This is
very much like to what happens in\textit{\ superstring theory}, in which (as
first recognized by Polchinski \cite{Po95}) superstrings end up on \textit{%
D-branes} Really, because of equations (\ref{col6}) and
(\ref{col7}), there is more than this. It is well known
\cite{VS00} that cosmic strings give rise to conical
singularities: roughly, this means that the geometry of surfaces
$(u,v=const)$ either for the metric (\ref{colright}) or for the
metric (\ref{coleft}) is not that of a plane, rather is that of a
cone. Moreover, the deficit angle $\Delta $ of these cones is
related in a simple way to the energy per unit of length $\mu $ of
the cosmic string:
\begin{equation*}
\Delta =8\pi G\mu .
\end{equation*}
Then, the two deficit angles of metrics (\ref{colright}) and (\ref{coleft})
are:
\begin{equation*}
\Delta ^{\pm }=8\pi G\sum_{i}\mu _{i}^{\pm }.
\end{equation*}
However, conditions (\ref{col6}) and (\ref{col7}) imply that the
two-dimensional gradients $\overrightarrow{\nabla }H^{\pm }$ (that is, the
gradients of $H^{+}$ and $H^{-}$ restricted to the ($x-y$)-cones orthogonal
to the propagation direction of the waves) coincide on $\Sigma $. Let $%
\gamma _{t}$ be a closed curve on $\Sigma $ encircling some of the
singularities in the ($x-y$)-cone at a given $t$. Because of Eqs (\ref{la1})
and (\ref{la2}) the two-dimensional Gauss law implies that
\begin{equation*}
\left\{
\begin{array}{c}
\left. \Delta \right| _{\gamma _{t}}^{+}\equiv \overset{N_{\gamma }}{\sum_{i}%
}\mu _{i}^{+}=\oint_{\gamma _{t}}\overrightarrow{\nabla }H^{+}\cdot
\overrightarrow{n} \\
\left. \Delta \right| _{\gamma _{t}}^{-}\equiv \overset{N_{\gamma }}{\sum_{i}%
}\mu _{i}^{-}=\oint_{\gamma _{t}}\overrightarrow{\nabla }H^{-}\cdot
\overrightarrow{n}
\end{array}
\right. ,
\end{equation*}
$\overrightarrow{n}$ being the external normal to $\gamma _{t}$ and $%
N_{\gamma }$ the number of singularities encircled by $\gamma _{t}$. Thus,
by taking into account Eqs (\ref{col6}) and (\ref{col7}), according to which
\begin{equation}
\oint_{\gamma _{t}}\overrightarrow{\nabla }H^{+}\cdot \overrightarrow{n}%
=\oint_{\gamma _{t}}\overrightarrow{\nabla }H^{-}\cdot \overrightarrow{n},
\label{coto}
\end{equation}
\  we conclude that
\begin{equation}
\left. \Delta \right| _{\gamma _{t}}^{+}=\left. \Delta \right| _{\gamma
_{t}}^{-}  \label{defiango}
\end{equation}
$\left. \Delta \right| _{\gamma _{t}}^{\pm }$ being the total deficit angle
encircled by $\gamma _{t}$; therefore, the deficit angles, the natural
topological charges occurring in this model, are conserved across the
membrane. Thus, since the above relation is topological in nature, \textit{a
posteriori} we can say that the use of the Israel conditions is an effective
tool also to study the collision. The above conservation law has also
another interesting consequence. At any given time $t$, the two kinds of
cosmic strings (the ones sourcing PP-waves travelling in the positive $z-$%
direction and the others sourcing PP-waves travelling in the negative $z-$%
direction) have to glue on $\Sigma $\ in the same points, otherwise it would
be possible to construct a $\gamma _{t}$ able to violate Eq. (\ref{coto})%
\footnote{%
This could be done by considering a $\gamma _{t}$ small enough to encircle
the intersection with $\Sigma $ of only one kind of cosmic strings, for
example, one on the left of $\Sigma $.}. Moreover, the previous argument can
be easily generalized in every dimension (it is fair to say that, in the
string theoretic literature, higher dimensional PP-waves coupled to
different kinds of p-forms already exist; we hope to study the relation
between the above model and superstring theory and, in particular, to extend
this simple model of scattering to superstring\ and $D$-branes theory in a
future work). In fact, PP-waves travelling in the positive $z-$direction can
always be written in the following form: with $A,B=5,..,N,$ and where $x$, $y
$ and the $w^{A\prime }s$ are \textit{transverse coordinates} and $N$ is the
dimension of the extended space-time. Then, it is trivial to see that
Einstein equations have the following PP-wave solution
\begin{equation}
g^{+}=2dudv+H^{+}(x,y,w^{5},..,w^{N},u)du^{2}-dx^{2}-dy^{2}-\delta
_{AB}dw^{A}dw^{B}\quad   \label{arbiPP}
\end{equation}
\begin{equation}
\left( \partial _{x}^{2}+\partial _{y}^{2}+\delta ^{AB}\partial _{A}\partial
_{B}\right) H^{+}=\sum_{i}\mu _{i}^{+}\delta ^{(N-2)}\left(
x_{i}(u),y_{i}(u),w^{5}(u),..,w^{N}(u)\right) ,  \label{poisson}
\end{equation}
where the sum runs on the spatial dimension along \ $\Sigma $; the same
argument also holds for a PP-wave travelling in the opposite spatial
direction. The Israel matching procedure also goes in the same way, the
final result being:
\begin{align*}
\partial _{i}H^{+}& =\partial _{i}H^{-},\quad i=x,y \\
\partial _{A}H^{+}& =\partial _{A}H^{-},\quad A=5,..,N.
\end{align*}

As one would expect, in this case also the Israel matching condition cannot
be fulfilled in vacuum but the introduction of a (mem)brane in $z=0$ does
the job. Of course, in this case also the deficit angle is conserved across $%
\Sigma$. In the next section we will analyze in some details the physical
meaning of the deficit angle in the five-dimensional case. Thus, cosmic
strings in $N$ dimensions join on an $N-1$\ dimensional (mem)brane (or, in
other words, on an $(N-2)$ branes). This raises the interesting possibility
to study a Kaluza-Klein-Randall-Sundrum scenario \cite{RS99a} \cite{RS99b}
with PP-waves travelling in the fifth dimension transversal to the brane and
to study their physical effects in the four ''standard'' dimensions where
cosmic strings would be perceived as a massive particles (since, with
respect to the induced metric on the brane, conical singularities follows a
timelike trajectory) generating, because of Eq. (\ref{poisson}), the
''Poissonian'' gravitational field of point-like particles in the $\left(
N-1\right) th$ dimensions of the brane while the radiative part of the
energy of the PP-waves travel in the ''$z$'' direction transversal to the
brane \cite{CVV02a} \cite{CVV02b}. Moreover, as already explained, such a
brane could not need any stabilization mechanism. Let us give a closer look
to the five dimensional case in which the matching hypersurface, in the
spirit of the brane-world scenarios \cite{HW96a} \cite{HW96b} \cite{Ar98a}
\cite{Ar98b} \cite{RS99a} \cite{RS99b}, can be interpreted as the
4-dimensional physical world.

\section{The 5-dimensional case}

To analyze the four-dimensional geometry of $\Sigma$ in the case in which
the space-time in Eq.(\ref{arbiPP}) is five-dimensional, it is useful to
introduce \textit{Minkowskian coordinates }$\left( t,z\right) $ by:
\begin{align*}
2t & =u+v \\
2z & =u-v,
\end{align*}
$z$ being the propagation direction of the PP-wave transversal to $\Sigma$.
In the new coordinates, the induced four-dimensional metric on $\Sigma $\
becomes:
\begin{equation}
\left. g\right| _{\Sigma}=(1+H)dt^{2}-dr^{2}-r^{2}d\Omega,  \label{RN1}
\end{equation}
where $r$, $\theta$ and $\varphi$\ are spherical coordinates and $H$ is a
solution of Eq. (\ref{poisson}) that, on $\Sigma$, reads
\begin{equation}
^{3}\Delta H=\sum_{i}^{q}\mu_{i}\delta^{(3)}\left( r_{i}(t),\theta
_{i}(t),\varphi_{i}(t)\right) ,  \label{poi1}
\end{equation}
where $^{3}\Delta$\ is the flat three-dimensional Laplace operator and the
positions of point-like sources may depend adiabatically on time\footnote{%
From the four-dimensional perspective, this means that if we move slowly
two, or many, point-like sources we do not make any work.}. It is easy to
see that this gravitational field corresponds to the weak field limit of $q$
spherically symmetric asymptotically flat black-holes, whose properties we
are going to discuss. The weak field limit is obtained by looking at the $%
g_{00}$ component of the metric. In our case, from the relation
\begin{equation}
g_{00}\sim1+2\Phi,  \label{pow1}
\end{equation}
we obtain for the Newtonian potential $\Phi$%
\begin{equation}
\Phi=\frac{1}{2}H.  \label{pow2}
\end{equation}

Thus, because of Eq. (\ref{poi1}), the metric asymptotically describes the
gravitational field of $q$ spherically symmetric black-holes of masses $%
\mu_{i}$. It is worth to note here that, since these $q$ spherically
symmetric black-holes are almost in equilibrium (that is, the relative
locations of the $q$ point-like sources in Eq. (\ref{poi1}) can be changed
slowly without doing any work), the four-dimensional metric in Eq.(\ref{RN1}%
) coincides, far away from the point-like sources, with the
Majumdar-Papapetrou solution \cite{Ma47} \cite{Pa47}. In fact, the
Majumdar-Papapetrou solution represents the gravitational field of $q$
spherically symmetric extremal charged black-holes in equilibrium because
the gravitational attractions is balanced by the electrostatic repulsion
\cite{Ch84}. More precisely, let us notice that a PP-wave in arbitrary
dimension can be sourced by an electromagnetic wave. In fact, if we
introduce in the space-time (\ref{arbiPP}), which in this section is assumed
to be $5$-dimensional, the \textit{potential }differential $1$-form $A$ and
the Faraday tensor field $F$
\begin{equation}
F=dA=du\wedge dH,  \label{ele1}
\end{equation}
with
\begin{align*}
A & =\widetilde{H}_{,i}dx^{i}, \\
\widetilde{H} & =\int H,
\end{align*}
then the associated energy-momentum tensor has only one non-vanishing
components $T_{uu}$, and can be represented as massless dust
\begin{equation*}
T_{\mu\nu}=\rho U_{\mu}U_{\nu},
\end{equation*}
$\rho$ being the energy density of the five-dimensional electromagnetic
field. The corresponding non vacuum $5$-dimensional Einstein equations read:
\begin{equation*}
^{3}\Delta H=\rho,
\end{equation*}
so that when $\rho$ approaches the finite sum\footnote{%
This means that we are considering the field generated by $q$ charged cosmic
strings which are in one-to-one correspondence with $\delta$-functions in
the r.h.s. of Eq. (\ref{poi1}).} of $\delta$-functions in the r.h.s. of Eq.(%
\ref{poisson}) we get back, on $\Sigma$, the gravitational field (\ref{RN1}%
). Restricted on the membrane $\Sigma$ the differential $2$-form field $F$
is purely electric
\begin{equation*}
\left. F\right| _{\Sigma}=dt\wedge dH,
\end{equation*}
and represents the Coulomb field of $q$ charged particles. To compute the
charges, say $\widetilde{\mu}_{i}$, let us write down the curved $5$%
-dimensional Maxwell equations:
\begin{equation*}
\nabla_{\mu}F^{\mu\alpha}=j^{\alpha},
\end{equation*}
$j^{\alpha}$ being the charge density five vector. The only non trivial
equation is
\begin{equation*}
\nabla_{\mu}F^{\mu\upsilon}=\sum_{i}^{q}\widetilde{\mu}_{i}\delta^{(3)}%
\left( r_{i}(u),\theta_{i}(u),\varphi_{i}(u)\right) \text{.}
\end{equation*}
Since the l.h.s. of the above equation is nothing but $^{3}\Delta H,$ the
charges are equal to the masses, because of Eqs. (\ref{poisson}) and (\ref
{poi1}). There is still another reason, related to the BPS states in
gravity, supporting this interesting relation between asymptotically flat
PP-waves and the Majumdar-Papapetrou solution. Indeed, in the case we are
dealing with, PP-waves admit a Killing spinor \cite{OW99} $\epsilon$ such
that
\begin{equation*}
\overline{\epsilon}\gamma^{\mu}\epsilon\partial_{\mu}=\partial_{v}=%
\partial_{t}+\partial_{z},
\end{equation*}
$\partial_{v}$ being the null Killing field of PP-waves ($z$ being the
spatial propagation direction of the PP-wave transversal to the membrane $%
\Sigma$) and $\gamma^{\mu}$\ the Dirac matrices. By restricting on $\Sigma$,
we get a space-time with a Killing spinor $\epsilon_{\Sigma}$ such that
\begin{equation}
\overline{\epsilon}_{\Sigma}\gamma_{\Sigma}^{\mu}\epsilon_{\Sigma}\partial_{%
\mu}=\partial_{t},  \label{gibbons}
\end{equation}
$\gamma_{\Sigma}^{\mu}$ being the $\gamma$-matrices restricted on $\Sigma$.
This is not a coincidence. Indeed, in order to have suitable stability
properties, BPS states in gravity should admit (see, for example, \cite{OW99}
\cite{GH82} \cite{GHHP83} \cite{GP84} and references therein) a Killing
spinor $\epsilon_{\Sigma}$ whose associated Killing vector is the timelike
Killing vector $\partial_{t}$ \footnote{%
That is, BPS states in gravity should obey Eq. (\ref{gibbons}).}. Moreover,
the Majumdar-Papapetrou solution (which, in fact, corresponds to a BPS state
in gravity \cite{GH82} \cite{GHHP83} \cite{GP84} \cite{OW99}) is the only
static solution of Einstein-Maxwell equations fulfilling Eq. (\ref{gibbons}%
). This is quite interesting because the BPS nature of the
Majumdar-Papapetrou solution, such as its stability, could be related to the
topology of charged cosmic strings. These results seem to be consistent with
previous results \cite{CM98}\footnote{%
We are indebted with prof. M. Bianchi for drawing this paper to our
attention.} obtained by analyzing D-brane dynamics in higher dimensions via
the Born-Infeld theory.

Finally, let us speculate upon an interesting feature of this model which is
related to the generalized Dirac quantization conditions in the brane-world
scenarios and, more generally, in superstring and D-branes theory and to its
BPS nature.

It is fair to say that, in the final part of this section, the mathematical
rigor partly will lack. However, we believe that an embedding of this model
in a superstring framework is possible, since results in \cite{CM98} in this
direction are rather encouraging. Such an embedding should strengthen the
following considerations.

Let us recall that, in the case of ''almost'' Majumdar-Papapetrou solution,
the BPS bound is given by:
\begin{equation}
M\geq\frac{1}{G}\sqrt{Q^{2}+\Pi^{2}},  \label{BPS2}
\end{equation}
where notation of \cite{OW99} have been used and where $G$ , $Q$ and $\Pi$
are the Newton constant, the electric charge and the magnetic charge
respectively. It would be nice to have a Dirac quantization condition that,
at least in the extremal case, is a quantization condition on the mass as
well. However, in the pure four dimensional case, quantization conditions
can only be achieved by going outside the framework of classical
supergravity with Abelian gauge fields \cite{OW99}. Interestingly enough, we
can get the desired relation by interpreting 4-dimensional
Majumdar-Papapetrou solutions as slices of\ a $5$-dimensional space-time
with superconducting cosmic strings, a sort of five dimensional analogue of
the Abrikosov-Nielsen-Olesen vortex solutions \cite{Ar57} \cite{NO73} (which
carry either electric or magnetic charges\footnote{%
Let us note that the five-dimensional electromagnetic field in Eq. (\ref
{ele1}) has either an electric or a magnetic component.}). In fact, for
superconducting cosmic strings the electric $\widetilde{Q}$ and magnetic $%
\widetilde{\Pi}$ charges satisfy the standard quantization condition (see,
for example, \cite{VS00})
\begin{equation*}
\widetilde{Q}\widetilde{\Pi}=2\pi n.
\end{equation*}
Since the electric and the magnetic charge, $\widetilde{Q}$ and $\widetilde
{\Pi}$, in the full space-time are nothing but the electric and magnetic
charges, $Q$ and $\Pi,$ on the slice $\Sigma$, we get, from Eq. (\ref{BPS2}%
),
\begin{equation}
M\geq\frac{\Pi}{G}\sqrt{1+\frac{4\pi^{2}n^{2}}{\Pi^{4}}}.  \label{BPS4}
\end{equation}
Then, in this model, the masses of extremal Reissner-Nordstrom black-holes
are quantized. Let us notice that $\Sigma$ can be regarded as the boundary
of a bulk in which there are propagating PP-waves; thus, as it could be
expected on holographic and string theoretic grounds, in the perturbative
spectrum of string theory, over PP-waves background, the $n^{th}$ quantized
mode (in the light cone gauge) of the string has energy \cite{BMN02}
\begin{equation}
\omega_{n}=\sqrt{\mu^{2}+\frac{n^{2}}{\left( \alpha^{\prime}p^{+}\right) ^{2}%
}},  \label{BPS5}
\end{equation}
$\alpha^{\prime}$ being the string tension and $\mu$ the ''strength'' of the
PP-wave.

The resemblance of Eqs. (\ref{BPS4}) and (\ref{BPS5}) could be a consequence
of the fact that, in many cases in which holography holds, the strong
coupling regime on the boundary (the BPS-extreme blackholes solutions on $%
\Sigma$) corresponds to the weak coupling regime of the bulk theory.
However, this sort of ''holographic'' duality really seems a coincidence. At
a first glance, from a stringy perspective, we are dealing with \textit{open
strings} in the bulk (that is, the one dimensional objects sourcing the
PP-wave) while the spectrum in Eq.(\ref{BPS5}) has been found by quantizing
\textit{closed strings} in the bulk \cite{BMN02} so that such superficial
duality should not hold\footnote{%
We thank M. Bianchi for pointing out this to us.}. In fact, this sort of
duality could hold anyhow. It is known \cite{Ch84} that the
Majumdar-Papapetrou solution is usually written in a coordinate system in
which non physical singularities appear at the positions $r_{i}=0$ where the
strings intersect $\Sigma$. At any fixed time $t$, the equations $r_{i}=0$
represent two-dimensional spheres, the areas of such surfaces being $%
\mu_{i}^{2}$ \cite{Ch84}. Thus, at any given time $t$, the intersections
between the strings and $\Sigma$\ are two-spheres. By taking into account
the transverse spatial dimension, we can say that the spatial shape is that
of the boundary of $S^{3}\times R$, that is a cylinder $S^{2}\times R$ (as
we will see in the next section, the internal space $S^{3}\times R$ bounded
by such a cylinder $S^{2}\times R$ can be ''filled'' with a quadratic
PP-wave which is well known to be related to closed strings). This
observation is not enough to restore the duality between the BPS spectrum on
$\Sigma$\ and the perturbative closed string spectrum in the bulk. In fact,
the bulk spectrum in Eq.(\ref{BPS5}) has been computed in a quadratic
PP-waves background, while the BPS spectrum of Eq. (\ref{BPS4}) has been
computed on the boundary of a background with singular PP-waves. Thus there
are two possibilities: or the resemblance of Eqs.(\ref{BPS4}) and (\ref{BPS5}%
) has no physical meaning at all, or there is a close relation between
singular and quadratic PP-waves.

\section{Quadratic PP-waves as the inner solution}

In the previous section it has been shown that the objects sourcing
asymptotically flat PP-waves are the boundaries of space-time regions whose
topology is $S^{3}\times R$ (in five space-time dimensions). In order for
the model to be reliable, a reasonable form for the internal metric has to
be found: interestingly enough, the internal metric turns out to be well
represented by quadratic PP-waves.

Now we will briefly describe the properties of quadratic PP-waves in
arbitrary dimensions. The metric representing a quadratic PP-wave
propagating in the positive $z-$direction is:
\begin{align}
g^{+} & =2dudv+H^{Q}(u,x^{i})du^{2}-\delta^{ij}dx^{i}dx^{j},  \label{QP1} \\
H^{Q} & =f_{ij}(u)x^{i}x^{j}.  \notag
\end{align}
However taking into account the vacuum field equations which imply
\begin{equation}
R_{\mu\nu}=0\Rightarrow Trf_{ij}=0,
\end{equation}
and the spherical symmetry in the transverse dimensions which implies
\begin{equation*}
f_{ij}=F(u)\delta_{ij},
\end{equation*}
we obtain $f\left( u\right) =0$ and $H^{Q}=0.$ It follows that a
spherically-symmetric quadratic PP-wave, which is the maximally symmetric
plane wave \cite{HR03}, cannot be a solution of the vacuum field equations
and has to be coupled with some (generalized) electromagnetic field. From
the previous discussion on the five-dimensional case, it follows that, near
each singularity in the transverse plane, singular PP-waves are spherically
symmetric and, moreover, are naturally coupled to higher-dimensional
electromagnetic fields vanishing outside the singularities. Thus, it is
natural to imagine that, inside the singularities, we have
spherically-symmetric quadratic PP-wave while outside we have singular
PP-waves. This is very much like to what happens in classical Newtonian
theory when we describe the gravitational field inside and outside a sphere
of constant density. To check this idea we have to apply the Israel's
procedure and to verify that the matching can be done in the simplest way as
possible. For definiteness, we will restrict ourself to the $5-$dimensional
case the generalization to higher dimensional cases being straightforward.
The matching hypersurfaces will be $\Sigma_{r_{i}^{\ast}}\equiv\left\{
\left. (u,v,r,\theta\varphi)\right| r=r_{i}^{\ast}\right\} $ where $%
r_{i}^{\ast}$\ is the position of the i-th singularity of $H^{+}$. To
perform the computation around the i-th singularity, we can place the i-th
singularity at the origin of the coordinates system, so that $K_{ab}$
becomes:
\begin{equation*}
K_{ab}-\frac{1}{2}h_{ab}=\left\{
\begin{array}{cccc}
0 & -1 & 0 & 0 \\
-1 & -\frac{\partial_{r}H}{2}+\frac{2H}{r} & 0 & 0 \\
0 & 0 & -r & 0 \\
0 & 0 & 0 & -r\sin^{2}\theta
\end{array}
\right\}
\end{equation*}
where $h_{ab}$ is the induced metric on $\Sigma_{r_{i}^{\ast}}$. We have to
fulfil Eqs.(\ref{col3}) and (\ref{col4}) using (\ref{QP1}) as the inner
metric, and (\ref{arbiPP}), in the $5-$dimensional case, as the outer
metric. In the general case (that is, when there are N singularities in the
outer solution), the matching equations (\ref{col3}) and (\ref{col4}) cannot
be solved in a closed form. However, if the singularities are far apart,
around each singularity we can neglect the influence of the others. Thus,
the matching equations read:
\begin{gather}
F(u)R_{M}^{2}=\frac{\mu_{i}(u)}{R_{M}}  \label{QP5} \\
F(u)R_{M}-\frac{3}{2}\frac{\mu_{i}(u)}{R_{M}^{2}}=S_{uu}  \label{QP6}
\end{gather}
where $R_{M}$\ is the matching radius. To have an idea of the order of
magnitude of $R_{M}$ and $S_{uu}$ the following considerations are useful.
From the ''outer point of view'', we have a gas of extreme blackholes, whose
horizons have areas proportional to $\mu_{i}^{2}$. Then, it is reasonable to
assume that the ''inner radius'' to be identified with $R_{M}$ is
proportional to $\mu_{i}$, so that $R_{M}$\ also would be quantized. From
Eqs. (\ref{QP5}) and (\ref{QP6}) we get
\begin{equation*}
F\sim\frac{1}{R_{M}^{2}},\quad S_{uu}\sim\frac{1}{R_{M}},
\end{equation*}
as expected from the Newtonian analogy. Eventually, we can say that the
Israel's procedure also works in this case. For these reasons, we can
naturally interpret quadratic PP-waves as the inner solutions of the
singular ones: that is, maximally symmetric quadratic PP-waves (as well as
their natural p-forms sources) live inside the singularities of the singular
PP-waves, very much like to what happens in the Newtonian theory of
spherical body. Thus, the resemblance of Eqs. (\ref{BPS4}) and (\ref{BPS5})
is not a coincidence, but it is likely to be a manifestation of a
holographic weak-strong duality. This result rises the interesting
possibility of analyzing the properties of space-times in which there are
zones with many SUSYs (that is, inside the singularities) and other zones
with less SUSYs (outside the singularities).

\section{Conclusion}

In this paper a model is proposed to study the matching between
asymptotically flat PP-waves in a $N$-dimensional space-time. The
hypersurface $\Sigma$\ can be thought as an ($N-1$)-dimensional timelike
hypersurface and the Israel matching conditions, constraining the
interactions, determine the conserved topological quantities across $\Sigma$%
. It turns out that the deficit angle, the topological charge related to
cosmic strings which are the natural source of PP-waves, is conserved across
$\Sigma$.

In the spirit of brane-world scenarios, we analyzed the $5-$dimensional case
in which $\Sigma$\ could be thought as our $4-$dimensional world. The $4-$%
dimensional geometry on $\Sigma$ corresponding to a $5-$dimensional
space-time with $q$ (charged) cosmic strings transversal to $\Sigma$
coincides asymptotically (that is, far away from the sources) with the
Majumdar-Papapetrou solution, a BPS state of gravity. Furthermore, although
in a $4-$dimensional context it is not possible to obtain any Dirac-like
quantization condition with only Abelian gauge fields without going outside
the framework of classical supergravity \cite{OW99}, by interpreting the $4-$%
dimensional Majumdar-Papapetrou solution in a $5-$dimensional perspective as
a slice of an asymptotically flat $5-$dimensional PP-waves solution with $q$
superconducting cosmic strings, we have got a Dirac quantization condition
and the related mass quantization for extremal blackhole in four dimension.
Such quantization, in a purely four-dimensional perspective, can be obtained
only by using supersymmetric QFT on the highly non trivial background
provided by the Majumdar-Papapetrou solution, not an easy task indeed. The
resulting BPS spectrum in four dimension seems to be dual to the
perturbative spectrum of closed strings on quadratic PP-waves background in
higher dimensions found in \cite{BMN02} \cite{Me02}. This interpretation has
been strengthened showing that the quadratic PP-waves can be naturally
interpreted as the inner solution of the singular ones. Of course, a genuine
superstring-brane theoretic interpretation of this model is needed to
provide sound physical basis to the previous result.

\begin{acknowledgement}
The authors wish to thank M. Bianchi, K. Panigrahi, A Sagnotti and O. Zapata
for interesting remarks and suggestions. One of them (G. V.) wishes to thank
Profs. P. Michor and J. Yngvason for the kind hospitality at the
Schr\"{o}odinger International Institute for Mathematical Physics in
Vienna.The work was supported in part by the Progetto di Ricerca di
Interesse Nazionale SINTESI\ 2002.
\end{acknowledgement}

\end{document}